\documentclass[10pt]{iopart}
\usepackage{epsf,epsfig,latexsym}
\usepackage{setstack}
\usepackage{amssymb}
\usepackage{bm}
\usepackage{graphicx}
\usepackage{dcolumn}
\usepackage{multirow}
\usepackage[rflt]{floatflt}
\usepackage{units}

\DeclareRobustCommand{\dgr}[1][]{
  \unit[#1]{\ifmmode{}^\circ\else${}^\circ$\fi}}

\def\ssNN#1{\sqrt{s_{NN}} \ifx|#1|\else=\unit[#1]{GeV}\fi}
\DeclareRobustCommand{\snn}[1]{\ifmmode\ssNN{#1}\else$\ssNN{#1}$\fi}
\DeclareRobustCommand{\valn}[3]{\ifmmode #1\,\pm\,_{#2\text{ (stat)}}^{#3\text{ (syst)}}\else$#1\,\pm\,_{#2\text{ (stat)}}^{#3\text{ (syst)}}$\fi}
\DeclareRobustCommand{\val}[3]{\ifmmode #1\,\pm\,_{#2}^{#3}\else$#1\,\pm\,_{#2}^{#3}$\fi}

\DeclareRobustCommand{\mpt}{\ifmmode\left<p_T\right>\else$\left<p_T\right>$\fi}
\DeclareRobustCommand{\pT}{\ifmmode p_T\else$p_T$\fi}
\DeclareRobustCommand{\mTm}{\ifmmode m_T-m_0\else$m_T-m_0$\fi}
\DeclareRobustCommand{\dy}{\ifmmode  \left<\delta y\right>\else $\left<\delta y\right>$\fi}
\DeclareRobustCommand{\my}{\ifmmode  \left<y\right>\else $\left<y\right>$\fi}

\newcommand{\nbi}           {$\rm^{a}$}

\begin{document}
\title[Charged Hadron Production with BRAHMS]{Rapidity
   Dependence of Charged Hadron Production in Central Au+Au
   Collisions at \snn{200} with BRAHMS}

\author{
  D.~Ouerdane\nbi for the BRAHMS Collaboration}
\address{\nbi~Niels Bohr Institute, University of Copenhagen, Denmark}

\begin{abstract}
  We have measured the rapidity distributions $dN/dy$ of $\pi^{\pm}$,
  $K^{\pm}$ and $p,\,\bar{p}$ in central Au+Au collisions at
  \snn{200}. The average rapidity loss per participant nucleon is
  $2.0\pm 0.2$ units of rapidity.  The strange to non--strange meson
  ratios $K/\pi$ are found to track variations of the baryo--chemical
  potential in energy and rapidity.
\end{abstract}

In ultra-relativistic heavy ion collisions, final state hadrons are
used as a probe of the different reaction stages, with a special focus
on observables that may reveal the existence of an early color
deconfined phase, the so--called quark gluon plasma.  The bulk
properties of the collision dynamics are becoming well understood at
mid--rapidity ($|y|\lesssim 0.5$).  However, in Au+Au collisions at
\snn{200}, where the beam rapidity is $y_{b} = 5.36$, this corresponds
to less than 10\% of the whole rapidity range ($2\times y_b$).  One of
the goals of the BRAHMS experiment~\cite{nim} is to explore a broader
rapidity range ($-0.1 \lesssim y \lesssim 3.5$). In this paper, we
report on some of our latest results on identified charged particle
yields from the 5\% most central Au+Au collisions at \snn{200}.\\

\begin{floatingfigure}[r]{7.9cm}
  \centering
  \includegraphics[width=0.48\columnwidth]{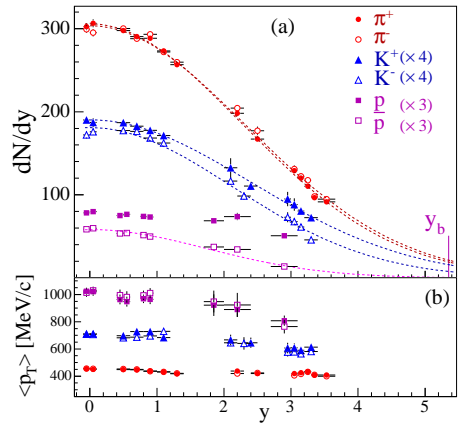}
  \caption{$\pi^\pm$, $K^\pm$ and $p,\bar{p}$ rapidity distributions
    (a) and mean transverse momentum \mpt{} (b). Errors are
    statistical.}
  \label{fig:dndy}
\end{floatingfigure}
BRAHMS consists of two hadron spectrometers, a mid--rapidity arm (MRS)
and a forward rapidity arm (FS), as well as a set of detectors for
global event characterization~\cite{nim}. Particle spectra were
obtained by combining data from several spectrometer settings
(magnetic field and angle), each of which covers a portion of the
phase--space $(y,p_T)$.  The data have been corrected for the limited
acceptance of the spectrometers using a Monte-Carlo calculation
simulating the geometry and tracking of the BRAHMS detector system.
Detector efficiency, multiple scattering and in--flight decay
corrections have been estimated using the same technique, but the data
have not been corrected for feed--down from resonance and hyperon
decays. Details can be found in~\cite{nim,myself}.

The pion and kaon spectra were well described at all rapidities by a
power law in $p_T$, $A (1 + p_T/p_0)^{-n}$, and an exponential in $m_T
- m_0$, $A \exp\left(\frac{m_T - m_0}{T}\right)$, respectively. The
invariant yields $dN/dy$ were calculated by integrating the fit
functions over the full $p_T$ or $m_T$ range. The systematic errors on
$dN/dy$, including errors from extrapolation and normalization, amount
to $\sim$ 10\% in the range $0 \lesssim y < 1.3$ and $\sim$ 15\% for
$y > 1.3$.  Rapidity densities and mean transverse momenta are shown
in Fig.~\ref{fig:dndy}.  Negative and positive pions are found in
nearly equal amounts within the rapidity range covered. In contrast,
an excess of $K^+$ ($p$) over $K^-$ ($\bar{p}$) is observed to
increase with rapidity. Figure~\ref{fig:dndy}(b) shows the rapidity
dependence of \mpt{}.  There is no significant difference between
positive and negative particles of a given mass.  In order to extract
full phase space densities for $\pi^\pm$ and $K^\pm$ we have
investigated several fit functions: a single Gaussian centered at
$y=0$ (G1), a sum of two Gaussians (G2) or Woods-Saxon (WS)
distributions placed symmetrically around $y=0$. In
Tab.~\ref{tab:4pimult} are reported the average yields.
\begin{table}[htb]
  \caption{Full phase--space yields of $\pi^\pm$ and $K^\pm$ extracted
    from fits to the $dN/dy$ distributions (see text). Errors 
    reported here are statistical, systematic errors are of the order of
    8\%.}  \centering
  \begin{tabular}{c@{\hspace{0.3cm}}c@{\hspace{0.3cm}}c@{\hspace{0.3cm}}c@{\hspace{0.3cm}}c}
         $\pi^+$       & $\pi^-$       & $K^+$       & $K^-$  &   $\bar{p}$\\\hline
        1742 $\pm$ 17 & 1761 $\pm$ 16 & 288 $\pm$ 5 & 241 $\pm$ 3 & 85
        $\pm$ 4
  \end{tabular}
  \label{tab:4pimult}
\end{table}

\begin{figure}[htb]
  \begin{minipage}[t]{.49\textwidth}
    \includegraphics[width=\columnwidth]{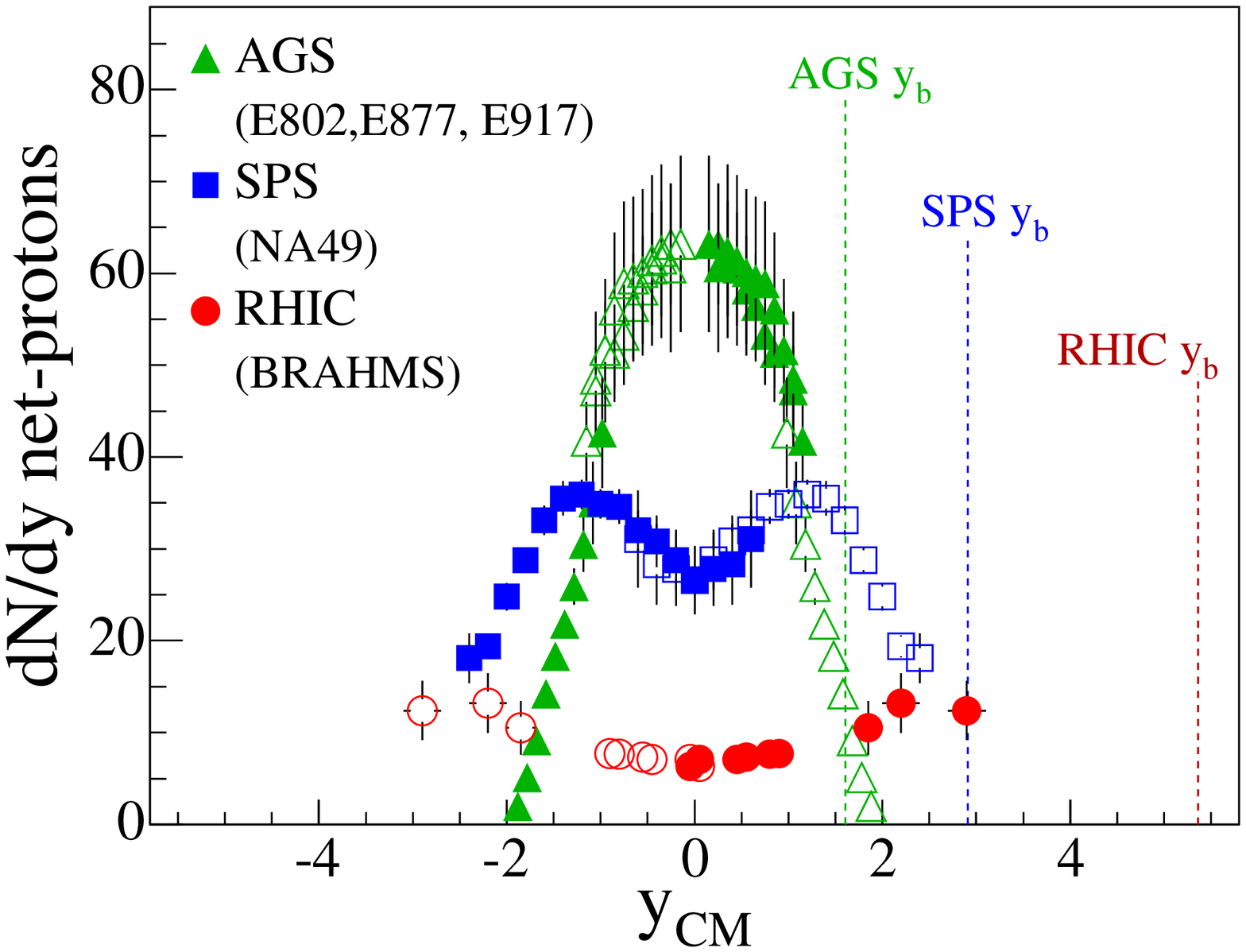}
    \caption{Net--proton rapidity densities as a function of rapidity
      for three different beam energies.}
    \label{fig:netp}
  \end{minipage}
  \begin{minipage}[t]{.49\textwidth}
    \includegraphics[width=\columnwidth]{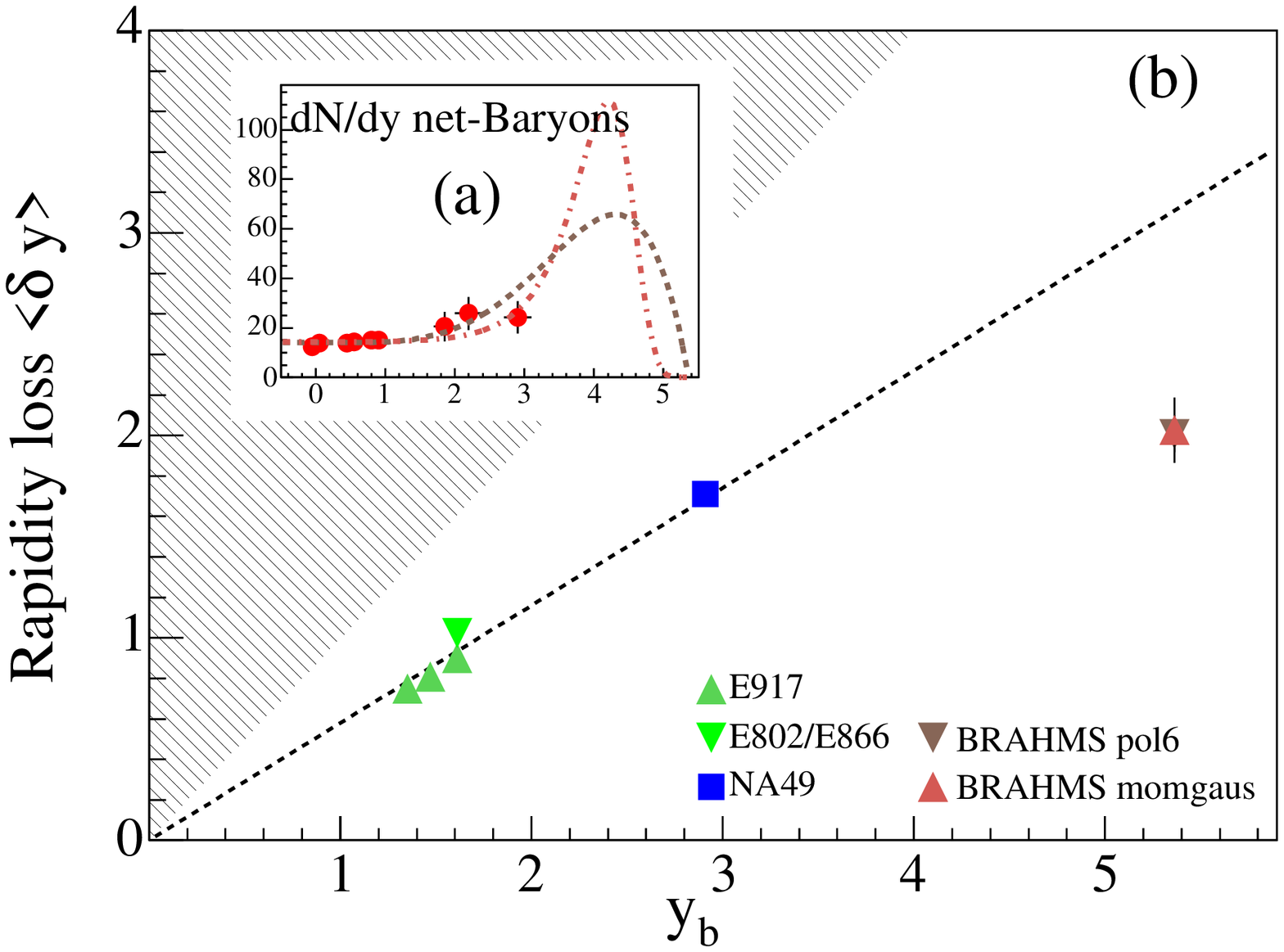}
    \caption{Net--baryon rapidity distributions (a)
      and average rapidity loss per participant nucleon as a function of
      beam rapidity (b). Errors are systematic.}
    \label{fig:stopping}
  \end{minipage}
\end{figure}

Figure~\ref{fig:netp} shows the net--proton densities $dN/dy(p) -
dN/dy(\bar{p})$ as a function of rapidity. Our
data~\cite{rhicnetp,peter} (red circles) are plotted together with
AGS~\cite{agsnetp1,agsnetp2,agsnetp3} and top SPS~\cite{spsnetp} data.
A drastic change in the shape of the distribution occurs as \snn{}
increases. At RHIC, the net--proton density is $6.4\pm 0.4\pm 1.0$
around mid--rapidity ($y < 1$) and increases to $12.4\pm 0.3\pm 3.2$
at $y\sim 3$. This indicates that collisions are much more transparent
than at AGS and SPS.  The energy available for particle production is
determined from the energy loss $\Delta E$ of the original nuclei and
is related to the degree of stopping. Due to the conservation of the
baryon number during the reaction, the final state net--baryon
($B-\bar{B}$) distribution contains information on $\Delta E$. The
nuclear stopping is quantified by the rapidity loss \dy{}, which, for
mass symmetric collisions, is defined by $\dy{} = y_b -
\left<y\right>$, where \my{} is the mean net--baryon rapidity after
the collision:
\begin{equation}
  \my{} = \int_0^{y_b} y\,\frac{dN_{(B-\bar{B})}(y)}{dy}\,dy
\end{equation}
It is therefore necessary to know the distribution of baryons in the
whole rapidity range. A Monte-Carlo simulation provided corrections
factors for the non--detected neutrons, and strange baryons such as
$\Lambda$'s and $\Sigma$'s. Only mid--rapidity hyperon yields are
known experimentally, so we made assumptions for more forward
rapidities. Details can be found in~\cite{rhicnetp,peter}.  Since the
data do not cover the entire rapidity range, the missing information
has been extrapolated from the measured data by fitting the latter
with a function that is continuous and whose integral is equal to the
number of participants ($357\pm 9$ for the 5\% most central collisions
after a Glauber simulation).  Figure~\ref{fig:stopping} shows the
net--baryon distribution as a function of $y$ (insert) and \dy{} as a
function $y_b$ (main panel).  Two functions were used: a sum of
Gaussians symmetrically positioned around mid--rapidity, and a six
order polynomial (dashed lines in insert). In both cases, \dy{}
amounts to $2.0\pm 0.2$. This result is significantly lower than the
expectation from the empirical scale $\dy{} \propto y_b$. The energy
loss, calculated after the net--baryon fits, is found to be $\Delta E
= \unit[72\pm 6]{GeV}$ per participant, i.e. $\unit[25.7\pm 2.1]{TeV}$
for the corresponding
centrality class.\\


\begin{figure*}[htb]
  \centering
  \includegraphics[width=0.9\textwidth]{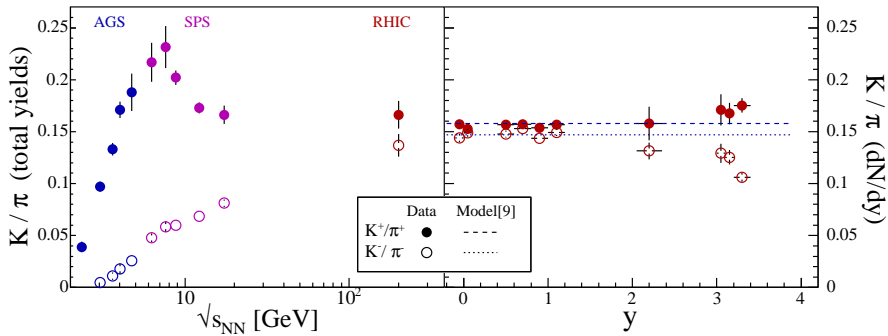}
  \caption{Full phase--space $K/\pi$ ratios as a function of \snn{}
    (a) and rapidity systematics at \snn{200} (b). The dashed and
    dotted lines in (b) are predictions of the statistical
    model~\cite{braun}. Errors are statistical and systematics in (a),
    only statistical in (b). AGS data are from~\cite{kpiAGS1,kpiAGS2},
    SPS data from~\cite{newna49,newna49b,kpiSPS1}. Data points at
    \snn{6.3} and \unit[7.6]{GeV}~\cite{newna49,newna49b} are preliminary.}
  \label{fig:kopi}
\end{figure*}
In Fig.~\ref{fig:kopi}(a) are shown ratios of the strange to
non--strange full phase--space meson yields ($K/\pi$) as a function of
\snn{}.  The ratio $K^+/\pi^+$ shows a fast increase from low AGS to
low SPS energies (\snn{8}), followed by a decrease with increasing SPS
energy (to \snn{17}). We find, at \snn{200}, a value of $0.165 \pm
0.003\pm 0.013$, consistent with the SPS ratio at the highest energy.
In contrast, $K^-/\pi^-$ increases monotonically but remains below
$K^+/\pi^+$.  At \snn{200}, it reaches a value of $0.136 \pm 0.002\pm
0.011$, which is close to $K^+/\pi^+$ at the same energy. The full
phase--space ratios are not significantly different from those
observed in the mid--rapidity region $y \lesssim 1$
(Fig.~\ref{fig:kopi}(b)). Indeed, a fit to a straight line in this
particular range gives $K^+/\pi^+ = 0.156\pm 0.023$ (stat + syst) and
$K^-/\pi^- = 0.146\pm 0.016$. The dashed and dotted lines are
predictions of the hadron gas statistical model~\cite{braun}, which
used a chemical freeze--out temperature $T$ of \unit[177]{MeV} and
baryo--chemical potential $\mu_B$ of \unit[29]{MeV} (the authors
restricted their fits to yields measured in the rapidity range
$|y|<0.5$). The agreement with the data is excellent. However, the
data deviate from the model prediction at higher rapidities, where an
increasing excess of $K^+$ over $K^-$ is observed. This is due to an
increase of net--baryon densities~\cite{rhicnetp,ratio200}. A baryon
rich environment is favorable for associated strangeness production,
e.g.  $p+p\rightarrow p+K^+ +\Lambda$, a production channel forbidden
to $K^-$.  In the context of the statistical model, this translates
into an increase of the baryo--chemical potential $\mu_B$, as already
reported in~\cite{ratio200}, where a calculation by Becattini {\it et
  al.}~\cite{becat} of $K^+/K^-$ vs $\bar{p}/p$ at constant $T$ (and
varying $\mu_B$) agrees well with the rapidity dependence of the
experimental ratios. It is also known that the chemical freeze--out
temperature varies strongly in the AGS energy range but slightly from
SPS to RHIC energies. In this particular energy domain, the
$K/\pi$ systematics are as well mainly driven by changes in $\mu_B$.\\

In summary, we have measured transverse momentum spectra and inclusive
invariant yields of charged hadrons $\pi^\pm$, $K^\pm$ and
$p,\bar{p}$. The net--proton densities are the lowest ever observed in
heavy--ion collisions and increase only slightly with increasing
rapidity. The rapidity loss calculated from the net--baryons is found
to be $\dy{} = 2.0 \pm 0.2$, thus breaking the empirical scaling
observed at lower energies.  The ratio of strange to non strange
mesons $K/\pi$ ratios are well reproduced by the hadron gas
statistical model~\cite{braun} that assumes strangeness equilibration.
The increasing difference between $K^+$ and $K^-$ yields at higher
rapidities is explained by a change of the baryo--chemical potential
$\mu_B$ with increasing rapidity.\\

This work was supported by the division of Nuclear Physics of the
Office of Science of the U.S. DOE, the Danish Natural Science Research
Council, the Research Council of Norway, the Polish State Com. for
Scientific Research and the Romanian Ministry of Research.

\section*{References}
\bibliographystyle{unsrt}
\bibliography{bibliography}

\end{document}